# Design and Simulation of a III-Nitride Light Emitting Transistor

Mohammad Awwad[1], Sheikh Ifatur Rahman[1], Chandan Joishi[1], Betty Lise Anderson[1] and Siddharth Rajan[1,2]

[1]Department of Electrical and Computer Engineering, The Ohio State University, Columbus, Ohio 43210, USA.
[2]Department of Materials Science and Engineering, The Ohio State University, Columbus, Ohio 43210, USA.

*Abstract*— **This paper describes the design and characteristics of monolithically integrated three-terminal gated III-Nitride light emitting diodes (LEDs) devices. The impact of channel doping and thickness on the voltage penalty of the transistor-LED hybrid device is analyzed, and it is shown that with appropriate design, low voltage drop can be realized across integrated gated LED structures. The impact of device design on the switching charge is investigated, and it is shown that the adoption of an integrated LED/transistor structure can reduce the switching charge necessary for operation of a switched LED display device by an order of magnitude when compared with stand-alone light-emitting diodes.**

*Index Terms*— **monolithic integration, microLED, MESFET, GaN.**

## I. Introduction

The III-nitride material system has been the key technology enabling solid-state lighting technologies, with widespread applications ranging from the field of general illumination to micro- light-emitting diode (LED) display systems and communications [1]–[5]. Emissive display technologies based on GaN/(In,Ga),N heterostructure light-emitting diodes with high pixel density, luminance, efficiency and large color gamut are of great interest for applications such as wearable technologies, mobile devices and virtual displays [6]–[8]. While high pixel density requirements for these display applications are fulfilled using scaled ($< 15 \ \mu m$) light-emitting diodes, scaling of the mesa dimensions lead to additional challenges for integration of LEDs and electronic drivers. Besides, it is desirable to have color and luminance variation created using a combination of three colors- red, green, and blue in micro-LED display technologies.

Given the performance variation in GaN/(In,Ga)N heterostructure-based LEDs at different emission wavelengths, it is necessary to use them at their optimum conditions to achieve maximum power savings. Extensive research on the integration of micro-LEDs for display applications in mobile devices, wearables and AR-VR technologies have shown great possibilities using both heterogeneous and monolithic integration [9]–[11]. The most widely used method to fabricate an active-matrix micro-LED display is to transfer the micro-LED chips onto a circuit board with a backplane [12], [13]. This method involves epitaxial growth of the LED wafer, patterning, etching, lift-off, and physical bonding of the micro-LED chips formed on different wafers that act as the sources of the red, green, and blue colors [14], [15]. This method performs well with large area displays with lower resolution but has performance and yield limitations when utilized in the fabrication of micro-LED displays for high-resolution and high-speed operation. Such challenges include difficulties in the alignment and bonding of the electrical interconnects of the micro-LED chips with the backplane circuitry, and a low production yield due to separate implementation of the three-color chips [16]. To solve these issues and introduce the next generation of display technology, power electronics combined with LEDs based on GaN are being pursued. Monolithic integration of GaN switching devices with the micro-LEDs sharing the same material platform offers certain advantages over heterogeneous integration. GaN transistors monolithically integrated on the LED for voltage driving would benefit from growth performed in a single epitaxial step and fewer fabrication processes, and the excellent performance offered by GaN transistors. The key advantage of a three-terminal gated LED is that it can greatly reduce the requirements on the switching devices driving LEDs since such a device enables *capacitive* rather than *current* control of the LED optical output. In case of the conventional two-terminal PN junction LEDs, the charge accumulated on the diode capacitance is relatively large since it is associated with charging and discharging the depletion region across the entire emitting active region. The circuits designed for switching these must supply this charge and the LED current. However, in the gated LED case, the device is switched by only modulating the gate charge ($Q_G$). Therefore, if the modulated gate charge is designed to be lower than the depletion region capacitance charge of the LED, the charge needed for switching the gated LED device is lower than an LED. Thus, the adoption of integrated 3-terminal gated LEDs can make control circuits simpler and more efficient.

Integration of GaN-based transistors and LEDs on the same epi-wafer can enable high power and high efficiency voltage-controlled modulated visible light emission while eliminating interconnects and reducing the spacing between each pixel. GaN based LEDs offer higher efficiencies of up to 80%, improved long-term reliability, and the ability to be manufactured at the microscale (less than 5 $\mu m$). Previous demonstrations of monolithic integration of GaN LEDs with GaN transistor technologies used regrown transistor structures on the side and interconnects between the LED and the transistors. Such integration methods possess challenges to device performance from the regrowth impurities and device spacing alongside a complicated process flow. Recently Hartensveld et. al demonstrated a voltage-controlled LED using a nanowire structure and a GaN FET for monolithic integration [17]. Bhardwaj *et al.* also showed light-emitting field-effect transistor (LEFET) using the benefits of bottom tunnel junction (TJ)-based LEDs [18]. In the report, they vertically integrated nanowires/Fin n-FETs with planar LEDs to control the LED emission profile. In this work, we discuss the design and characteristics of *monolithic planar* gated LED structures that are well-suited for existing LED process flows. The proposed design has a much lower switching charge compared to the stand-alone LED and removes the requirement for the driver circuit to provide LED drive current and stored charge. Therefore, using this design, we aim to reduce the current requirements and complexity of the controlling circuits, and to enable future intelligent multifunctional displays.

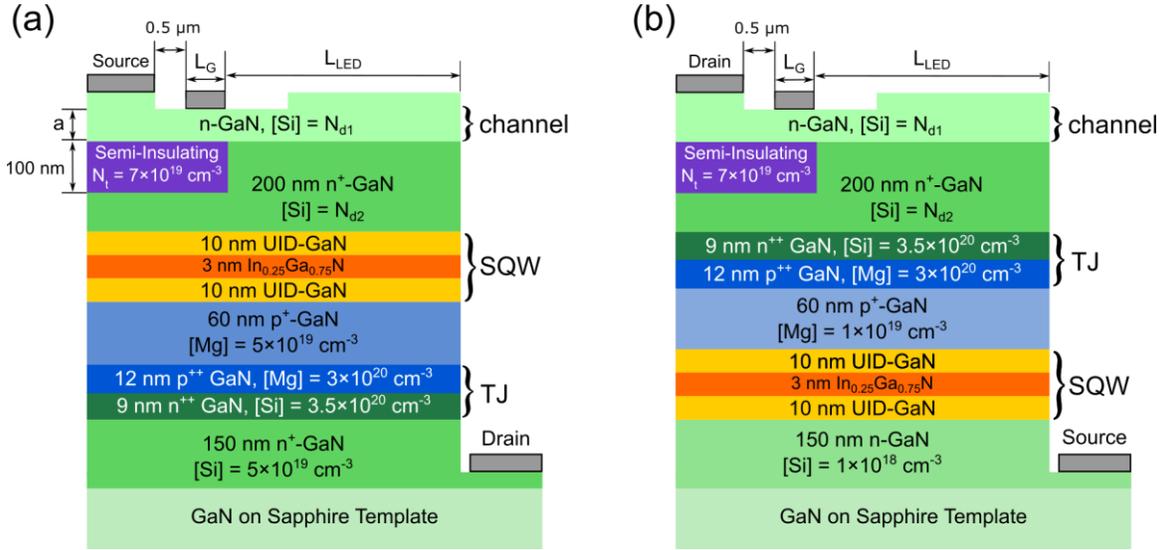

**Figure 1.** Two-dimensional schematics of the Light Emitting Transistors (LETs): (a) P-down configuration. (b) P-up configuration

## II. THEORY AND DESIGN

The schematics of the device designs investigated here is shown in Figure 1. It consists of a metal-semiconductor field-effect transistor (MESFET) with GaN/(In, Ga)N based heterostructure LEDs. For the p-down structure, the light emitting transistor (LET) is designed using a bottom tunnel junction-based structure as shown in fig. 1(a), which enables a conductive n-type region on the top surface, which can then be used to create a conductive channel for an n-channel transistor. P-down LEDs (using bottom tunnel junctions) have been reported with low voltage operation [19] and low injection barrier for carriers into the active region [20]. We note that the analysis here is also relevant for both p-up devices with a top tunnel junction as shown in fig. 1(b), with some differences that will be discussed in this manuscript. Recently, experimental demonstration of such an LET was achieved, and is reported separately.

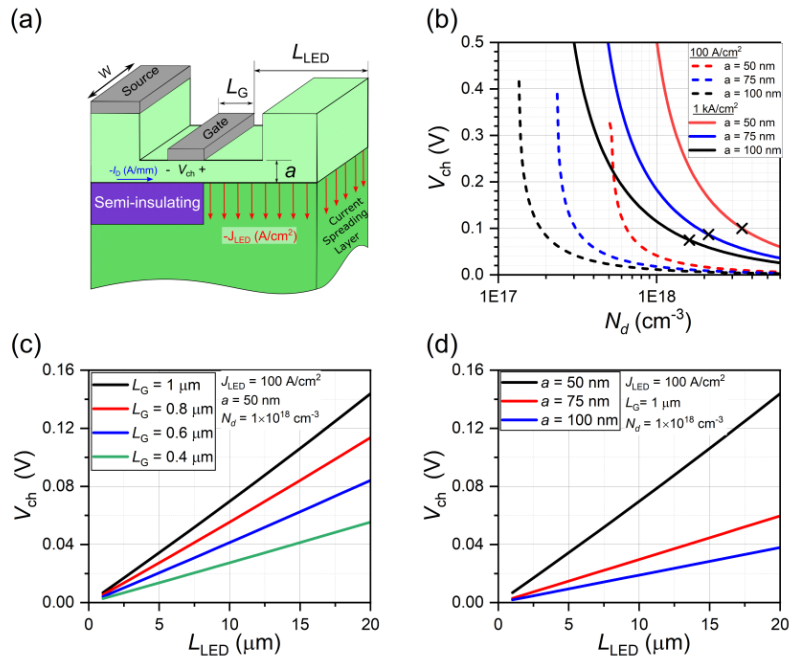

**Figure 2.** (a) 3-dimensional schematic of the MESFET with labelled parameters. (b) Channel voltage versus doping concentration at various channel thicknesses and currents, x markers indicate doping levels above which channel breaks down at full depletion. (c) Channel voltage versus $L_{LED}$ at various gate lengths. (d) Channel voltage versus $L_{LED}$ at various channel thicknesses.

The simulated device consists of a p-down $In_{0.25}Ga_{0.75}N$ QW LED on the top of a GaN homojunction TJ and a lateral depletion-mode MESFET transistor integrated at the top. The fully transparent homojunction TJ design has already been demonstrated to show low voltage drop for LED operation [21], [22]with a reduced voltage penalty. A green emitting (520 nm) LED above the homojunction TJ was implemented through a 3 nm $In_{0.25}Ga_{0.75}N$ single quantum well (SQW) sandwiched between two undoped GaN barrier layers. No additional AlGaN blocking layer was required in this design for the LED due to a built-in barrier in the InGaN QW that arises from the polarization charges [20].

A GaN MESFET was added above the LED on an n-doped channel layer, formed at the top of the heavily doped n-type current spreading layer. Vertical electron blocking from the source of the transistor was realized in the design with a semi-insulating (SI) blocking layer. Such a blocking layer profile below the source and gate could be achieved by introducing acceptor-like traps via ion implantation, typically done using nitrogen or helium ions [23], with a density larger than the background donor concentration. This would create a vertical electron-blocking path from the source to the drain but permit electron flow through a narrow lateral channel. Optimal values of doping density and thickness of n+-GaN layer below the channel were considered during the simulation to ensure efficient current spreading in the device. Source and drain contacts were assumed to be perfectly ohmic in this simulation, while the gate Schottky contact with a metal work function of 5.15 eV was considered [24], which results in a barrier height ~1 eV. A recessed gate design enabled high channel-length-to-thickness ($L_G/a$) aspect ratio without compromising the source resistance. The design parameters for the simulation including gate length ($L_G$), channel thickness ($a$), channel doping level ($N_d$), and LED length ($L_{LED}$) were optimized for high on-off LED current ratio, strong gate control, and reduced gate charge ($Q_G$) to promote fast switching.

Our target is to achieve a high on-off current ratio while still maintaining low channel resistance. We consider device designs that enable LED current ratings of up to 1 kA/cm$^2$ in the on-state, while simultaneously having strong gate control to suppress current < 0.01 A/cm$^2$ in the off-state. To determine the channel parameters for the LED current requirements, we used the gradual channel approximation (GCA) mathematical model for a MESFET [25], [26]:

$$I_D = I_{max}\left[\frac{V_{Ch} - V_{GS}}{\phi_{00}} - \frac{2}{3}\left(\frac{\phi_{bi} - V_{GS} + V_{Ch}}{\phi_{00}}\right)^{3/2} + \frac{2}{3}\left(\frac{\phi_{bi} - V_{GS}}{\phi_{00}}\right)^{3/2}\right] \tag{1}$$

such that:

$$I_{max} = \frac{qaWN_d\mu_n\phi_{00}}{L_G}, \qquad W = \frac{I_D}{J_{LED}L_{LED}}. \tag{2}$$

where $q$ is the elementary charge, $\mu_n$ is the doping dependent electron mobility in the channel estimated by the Albrecht model [27], $\phi_{00}$ is the full depletion potential, $V_{Ch}$ is the voltage across the channel ($V_{DS}$ in a conventional transistor), $W$ is the device width, and $J_{LED}$ is the current density passing through the LED in A/cm$^2$ whose length is $L_{LED}$. A 3-dimensional schematic of the MESFET part with labelled parameters is shown in Fig. 2(a). From (1) and (2), we can get an estimate of the voltage penalty $V_{Ch}$ added by the transistor at a normally-on condition ($V_{GS} = 0$) gate bias as shown in fig. 2(b). The maximum channel thickness $a_{max}$ is limited by the breakdown field $F_{BR}$ of GaN at full depletion such that: $a_{max} = {F_{BR}\epsilon}/{qN_d}$. The practical channel thicknesses considered here are still below $a_{max}$ for a wide range of doping levels owing to the relatively high breakdown field of GaN (3.3 MV/cm) [28]. From the plot, we observe that decreasing the channel doping below $10^{18}$ cm$^{-3}$ will significantly increase the voltage penalty at high injection regime for channel thickness, $a = 50$ nm. However, increasing the channel doping beyond $3 \times 10^{18}$ cm$^{-3}$ would increase the electric field near the metal-semiconductor Schottky junction at pinch-off as shown by the marked breakdown points in fig. 2(b). Therefore, a channel doping concentration of $10^{18}$ cm$^{-3}$ and thickness of 50 nm was adopted to achieve low channel resistance and strong gate control for small area micro-LEDs ($L_{LED} < 5\mu m$).

Micro-LEDs with larger area ($5\ \mu m < L_{LED} < 20\ \mu m$) require higher transistor current ratings ($I_D$) to maintain $J_{LED}$. Increasing $I_D$ with the same channel resistance results in increasing the channel voltage ($V_{Ch}$). Figures 2(c) and 2(d) show a plot of the calculated $V_{Ch}$ versus $L_{LED}$ for various gate lengths ($L_G$) and channel thicknesses ($a$), respectively. From the two plots, we observe an increase in the device voltage penalty as the micro-LED area increases. Therefore, reducing the channel resistance via shorter $L_G$ and/or higher channel thickness is necessary for larger area micro-LEDs.

## III. SIMULATION AND RESULTS

To validate the feasibility of the proposed design, the DC characteristics, optical power spectrum, and AC characteristics were simulated using the Silvaco simulation tool. Physics models (listed in Table I) were selected to match the processes taking place in each region. We used a SQW instead of a MQW region in the design and simulation to reduce the complexity and reach convergence faster, since the design is considered as an integrated multi-device design. Spontaneous and piezoelectric polarization interface charge were calculated using previous models [29]–[31] and included in the simulation.

TABLE I
PHYSICS MODELS USED IN THE SIMULATION

| Model | Description |
|---|---|
| Fermi-Dirac | Statistics of carrier concentration in doped regions |
| Incomplete ionization | A model that accounts for dopant freeze-out. Typically, it is used at low temperatures. |
| Albrecht | Low Field mobility model |
| k.p | Kronnig-Penney model effective masses and band edge energies for drift-diffusion simulation. |
| Thermionic emission | A model for the transport of energetic carriers across a semiconductor-semiconductor interface |
| Non-local band-to-band tunneling | Allows modeling of tunneling current between heavily doped pn regions |
| Concentration-dependent lifetime | Shockley-Read-Hall recombination using concentration dependent lifetimes |
| Spontaneous recombination | computes total radiative recombination rate and include it into drift-diffusion equations |
| Optical generation/Radiative recombination | Quantitative model for electron-hole pair optical generation/ radiative recombination process. |

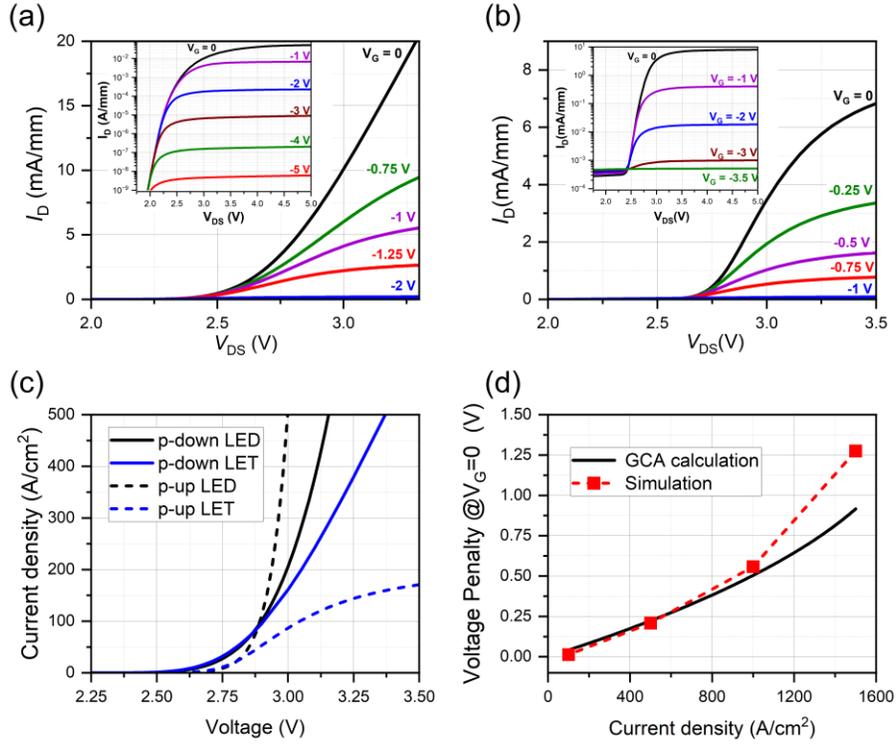

**Figure 3.** (a, b) Simulated linear I$_D$-V$_{DS}$ characteristics from V$_G$ = 0 to V$_G$ = -5 V, inset: logscale $I_D$-$V_{DS}$ from V$_G$ = 0 to V$_G$ = -5 V for (p-down, p-up) LET (c) Overlayed I-V characteristics of both proposed LET devices and the reference TJ LEDs. (d) Simulation vs analytical model comparison of voltage penalty (excess voltage) due to the addition of a controlling transistor.

Fig. 3(a) shows the $I_D$-$V_{DS}$ characteristics of p-down LET for $V_{GS}$ = 0 to -5 V for a device with $L_G = 1\ \mu m$, and $L_{LED} = 4\ \mu m$. Other I-V simulation results for $L_{LED} = 10\ \mu m$ and $L_{LED} = 20\ \mu m$ (not shown here) coincides with that shown in Fig. 3(a). As shown in the plot, the device turns on at ~ 2.5 V, which is similar to the turn-on voltage of the TJ LED device without the transistor. High on-off current ratios of ~$8 \times 10^5$ at V$_{DS}$ = 2.9 V ($J_{LED} = 100$ A/cm$^2$) and ~$7 \times 10^6$ at V$_{DS}$ = 3.95 V ($J_{LED} = 1$ kA/cm$^2$) are observed. The equivalent $I_D$-$V_{DS}$ characteristics of p-up LET are shown in fig. 3(b). A turn-on voltage of ~ 2.7 V can be observed which is higher than that of p-down LET due to the higher injection barrier as mentioned earlier. This effect is expected to be amplified on using MQW instead of SQW. To study the voltage penalty introduced by the controlling MESFET, a reference TJ LED is simulated with exactly the same design parameters as the proposed device except that both the implant-insulated region and the gate contact were removed. Fig. 3(c) shows the *I-V* characteristics of the proposed LETs and the reference stand-alone TJ LED previously mentioned. For the p-down LET, the integration of the gated structure to the standard p-down LED does not impact the diode performance up to the current density of 100 A/cm$^2$. At higher current densities, the device shows increased voltage drop due to the channel resistance. However, for the p-up LET, the excess voltage introduced by the transistor dramatically increases at current densities above 30 A/cm$^2$. This can be attributed to the proximity of the gate to the drain contact in p-up structure which resulted in channel pinch-off affected by the drain bias at low gate voltage. The excess voltage values at various current densities were extracted and are plotted versus current density for p-down LET in Fig. 3(d). For better understanding and validation, analytical calculations using the gradual channel approximation were also done. As shown in Fig. 3(c), a good agreement was found between GCA-calculated excess voltage and simulated excess voltage at low (near zero at $J_{LED}$ <100 A/cm²) and moderate injection (0.2 V at 500 A/cm²) regime. At high injection (>1 kA/cm²), however, 2D device simulations predict a voltage penalty that is higher than that calculated from GCA.

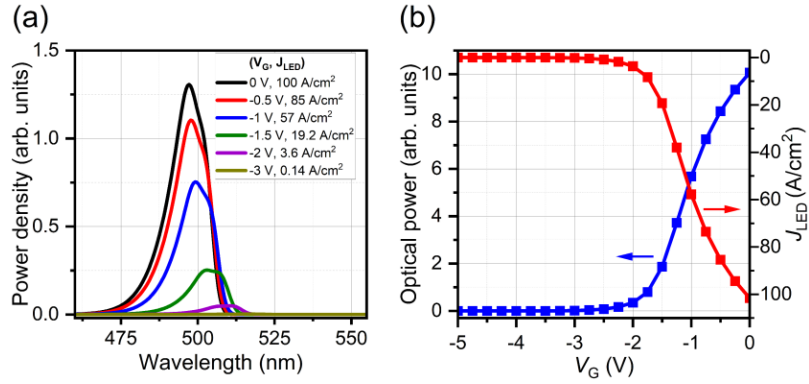

**Figure 4.** (a) Simulated emission spectra at $V_{DS}$ = 2.9 V (corresponding to 100 A/cm² at $V_{GS}$ = 0). (b) Optical output power and gate current versus gate voltage at $V_{DS}$ = 2.9 V.

The extra voltage penalty at high injection is due to current crowding near the drain side of the channel which can be solved by increasing the doping level and/or the thickness of the current spreading layer above the active emission area of the LED.

Fig. 4(a) shows the generated optical power spectral density as a function of the wavelength at $V_{DS}$ = 2.9 V which drives a current density of 100 A/cm² in the on state ($V_{GS}$ = 0). A blue shift is observed in high injection due to quantum-confined Stark effect (QCSE) which is common in InGaN QWs due to the applied electric field [32]. Total optical output power versus gate voltage shown in Fig. 4(b) is extracted by integrating the power spectrum in Fig. 4(a). When switching off the device, the emitted optical power is suppressed by a factor of ($\sim 1 \times 10^6$), which reflects the possibility of a true dimmed device in the off state and high dynamic range.

Under dynamic operation, the rise/fall time delays can be estimated by calculated the charge density accumulated/removed at the input terminal. An AC simulation was done in Silvaco to estimate the gate charge from the gate capacitance. Figure 5(a) shows the small-signal gate capacitance as a function of gate bias from $V_{gs}$ = 0 V to $V_{gs}$ = -5 V (giving $\sim 10^5$ on/off ratio). To calculate the gate charge, we integrated the gate capacitance over the switching range of $V_{gs}$ at various gate lengths. Fig. 5(b) shows $Q_G$ as a function of $L_G$, with the gate voltage varied such that the current decreases from 10 A/cm$^2$ to ~0. The simulated gate charge was estimated to be 7.2 pC/mm for $L_G = 1$ µm and 3.2 pC/mm for $L_G = 0.4$ µm. In comparison, the gate charge of the reference LED was estimated to be 27.97 pC/mm. This shows that the charge necessary to switch the gated LED is significantly lower than that needed for a stand-along LED.

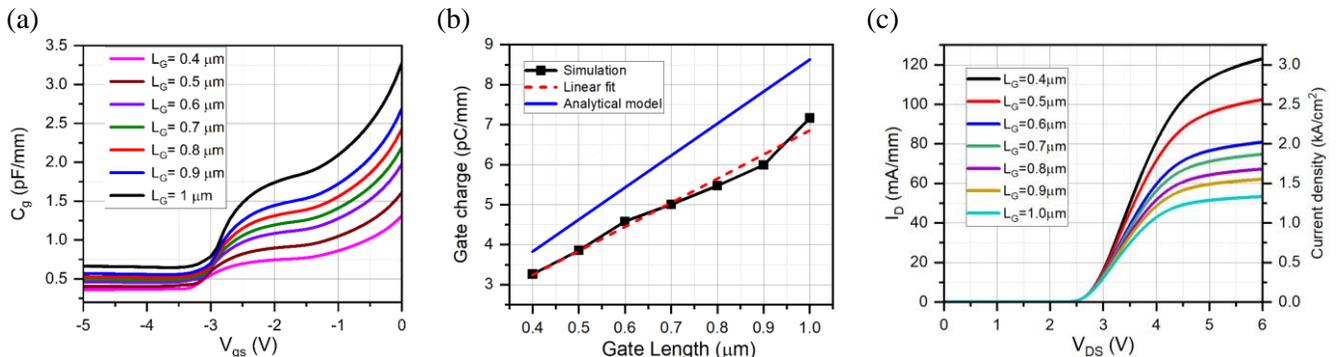

**Figure 5.** (a) Small-signal gate capacitance as a function of gate bias. (b) Extracted gate charge ($Q_G$) versus gate length ($L_G$) at 10 A/cm² (c) Simulated $I_D$-$V_{DS}$ curves at $V_{GS}$ = 0 V for various gate lengths ($L_G$).

The results agree with the derived analytical formula:

$$Q_G = \frac{\pi a^2}{2}(qN_d) + qN_d a\, L_G \tag{3}$$

where the first term represents the cylindrical charge at both edges of the gate contact, and the second term represents the charge under the gate. However, the model over-estimates the charge since, in reality, the depletion in the lateral direction is not as wide as in the normal direction which results in a non-perfect cylindrical depletion profile.

Reducing the gate length can enhance the switching characteristics of the device, however, the gate control over the channel might be affected by that reduction in the channel aspect ratio. Therefore, $I_D$-$V_{DS}$ simulations were repeated for the device at various gate lengths from 0.4 µm up to 1 µm as shown in Fig. 5(c). The on current (at $V_{GS} = 0$ V) increases with decreasing $L_G$ due to the reduction in channel resistance. In other words, at 100 A/cm$^2$ the device voltage is decreased from 2.9 V to 2.78 V by reducing the gate length from 1 µm to 0.4 µm, while the on-off current ratio is slightly reduced from ~$8 \times 10^5$ to ~$4 \times 10^5$. This shows that shorter gate lengths can provide sufficient on/off ratio while providing almost ~ 9X reduction in the gate switching charge.

## IV. SUMMARY

To conclude, we designed a monolithically integrated MESFET controlled tunnel-injected 520 nm LED for micro-LED display systems. The gate of the MESFET designed here controls the electron flow in the LED active region which in turn controls the emission profile of the LED. The design enables high power and efficient voltage modulation of LED with a single epitaxial step. Simulation results show a high on-off current ratio (>$10^5$) at 100 A/cm$^2$ and (>$10^6$) at 1 kA/cm$^2$ for high-dynamic-range displays. We also showed that the transistor introduces near zero voltage penalty at 100 A/cm$^2$ indicating a highly efficient operation. The low off current (<1 mA/cm$^2$) allows pixels to feature a true black state. The switching charge for an integrated transistor LED proposed here is shown to be significantly lower than a stand-alone LED. We note here that while the 3-terminal device does introduce excess voltage over a stand-along LED, it can greatly reduce the complexity of control circuits needed in micro-LED display applications.